\documentclass[conference]{IEEEtran}
\ifCLASSINFOpdf
\else
\fi
\usepackage{epsfig,mathptmx,times,amsmath,color,url}

\begin{document}
%

\title{
Learning Price-Elasticity of Smart Consumers in Power Distribution
Systems
}

\author{\large Vicen\c{c} G\'omez$^{1}$, Michael Chertkov$^{2}$, Scott Backhaus$^{3}$ and Hilbert J. Kappen$^{1}$\\
\small $^1$ Radboud University Nijmegen, The Netherlands. Donders Institute for Brain, Cognition and Behaviour, Nijmegen, The Netherlands\\
\small $^2$ Theory Division \& Center for Nonlinear Studies at LANL, Los Alamos, NM 87545\\
\small $^3$ MPA Division at LANL, Los Alamos, NM 87545
}
\maketitle

\begin{abstract}
Demand Response is an emerging technology which will transform the power grid
of tomorrow. It is revolutionary,  not only because it will enable peak load
shaving and will add resources to manage large distribution systems, but mainly
because it will tap into an almost unexplored and extremely powerful pool of
resources comprised of many small individual consumers on distribution grids.
However, to utilize these resources effectively, the methods used to engage
these resources must yield accurate and reliable control.  A diversity of
methods have been proposed to engage these new resources.  As opposed to direct
load control, many methods rely on  consumers and/or loads responding to
exogenous signals, typically in the form of energy pricing, originating from
the utility or system operator.
Here, we propose an open loop communication-lite method for estimating the
price elasticity of many customers comprising a distribution system. We utilize
a sparse linear regression method that relies on operator-controlled,
inhomogeneous minor price variations, which will be fair to all the consumers.
Our numerical experiments show that reliable estimation of individual and thus
aggregated instantaneous elasticities is possible. We describe the limits of
the reliable reconstruction as functions of the three key parameters of the
system: (i) ratio of the number of communication slots (time units) per number
of engaged consumers; (ii) level of sparsity (in consumer response); and (iii)
signal-to-noise ratio.
\end{abstract}


%
\IEEEpeerreviewmaketitle

\section{Introduction}
\label{sec:intro}
Today's Demand Response (DR) focuses on controlling major commercial and
industrial loads, i.e. large individual loads, where the actual control is infrequent and mostly focused on shaving peaks during times when the transmission grid and generation resources
are highly stressed \cite{DOEDR}.  Large peaking events are usually predicted
well in advance so that communication requirements for this type of DR duty are
quite limited; often taking the form of phone calls \cite{ENERNOC,DOEDR}.  At
other times, this large-scale DR may be used as a type of spinning reserve to
rebalance generation and load after a major grid disruption
\cite{ERCOTWind,PJMmarket}.  In this case, the immediacy of the need for the
resource justifies the cost of installing the communication so that the load
interruption is under direct control of the system operator.

As utilities and system operators integrate more time-intermittent renewables,
they will also be forced into a situation where there is less traditional
controllable generation resources online as there will be less room left in the
generation stack for these resources.  The loss of controllable resources will
occur at a time when they are needed even more to balance the intermittent
renewables.  Increased deployment of the DR is expected to be one controllable
resource that will fill this gap \cite{DOEDR}, however, the type of resource
required for this duty is different than the large-load DR discussed above.
Perhaps the most significant differences are that (a) this new form of DR will
be called upon more frequently, and (b) the control will be required to both
decrease and increase in a controlled fashion the load.

Accessing DR at the residential scale can be done via arrangements similar to
those currently used for large commercial and industrial customers, e.g.
contracts where customers receive payments or lower energy rates for providing
DR services.  However, it is expected that the majority of residential
consumers would balk at the idea of a utility or system operator have direct
control over loads within their home.  Instead, it is expected that DR will be
implemented via variable pricing or some other similar signaling \cite{DOEDR}.
Several models exist for this type of DR control, and they can be categorized
into two fundamental groups: open loop or closed loop control.  Retail-level,
double auction markets (also termed ``transactive control'') \cite{OlyPen}
represent one type of the closed loop control.  In this model, the control loop
is closed via a forward energy market where the supplier and each consumer
agree upon the amount of energy each load will consume and the price of energy
over the next market period.  Advantages of this type of control include
certainty about the energy consumption over the following market period and the
ability to build in network and/or generation constraints into the control in a
logical manner, e.g via local marginal pricing.  A significant drawback of this
type of control is the need for two-way, individually addressed communication
between the utility or system operator and every individual participating load.
The communication is not required to be real-time, however, the gathering of
energy bids from the loads must take place every market period which can be as
short as every five minutes.  Mechanisms other than double auctions have been
proposed to settle on energy quantity and pricing \cite{11CH}, however, the
two-way communication infrastructure and overhead remain essentially the same.

An alternative to the transactive control is open loop control where the
utility or system operator simply broadcasts a price to all participating
loads. The communication in this case is a simple one-way broadcast that does
not require any information to be returned from the customer--a form of
communication that is easier and less expensive to implement and that also does
not expose sensitive consumer data in a real-time environment.  Prices may be
updated on regular intervals with allowances for unscheduled updates triggered
by system disruptions.  After receiving an updated price, each participating
load consumes electricity at the current price if it desires
\cite{10TSBC,11TBAC}, however, the simplicity of the communication systems comes at a
cost of not having certainty about load response that the price change will
elicit.

In this work, our goal is to develop and demonstrate algorithms that
reduce the load response uncertainty in open loop control methods by estimating or learning the future price elasticity of consumers based on their responses to previous pricing updates.
We seek to keep communication requirements at a minimum raising a
significant challenge--how can we learn the price elasticities of {\it
individual} consumers and/or loads without deployment of additional sensors in
the distribution network and without resorting to two-way communication?  By
limiting our algorithms to sensing of power flows at the beginning of a
distribution circuit (where there is typically a sensor already installed), we
must resort to another method to distinguish individuals.  To solve the problem,
we consider multi-cast communication where we are able to address prices to
individual customers. We propose to introduce fluctuations in the individual
prices of each customer to enable estimating their individual
price elasticities. We express the task of learning the elasticities as a
linear regression problem \cite{Hoerl1,frank93,lasso,relaxed,vg,Friedman} in
which the aggregated changes in consumption over the distribution network are
represented as the weighted sum of all individual changes in consumptions.  The
prices enter in the model via the design matrix, and thus can be considered as
controlled variables chosen in a convenient way for the task under
consideration.

\begin{figure}[t]
      \centering
      \includegraphics[scale=.6]{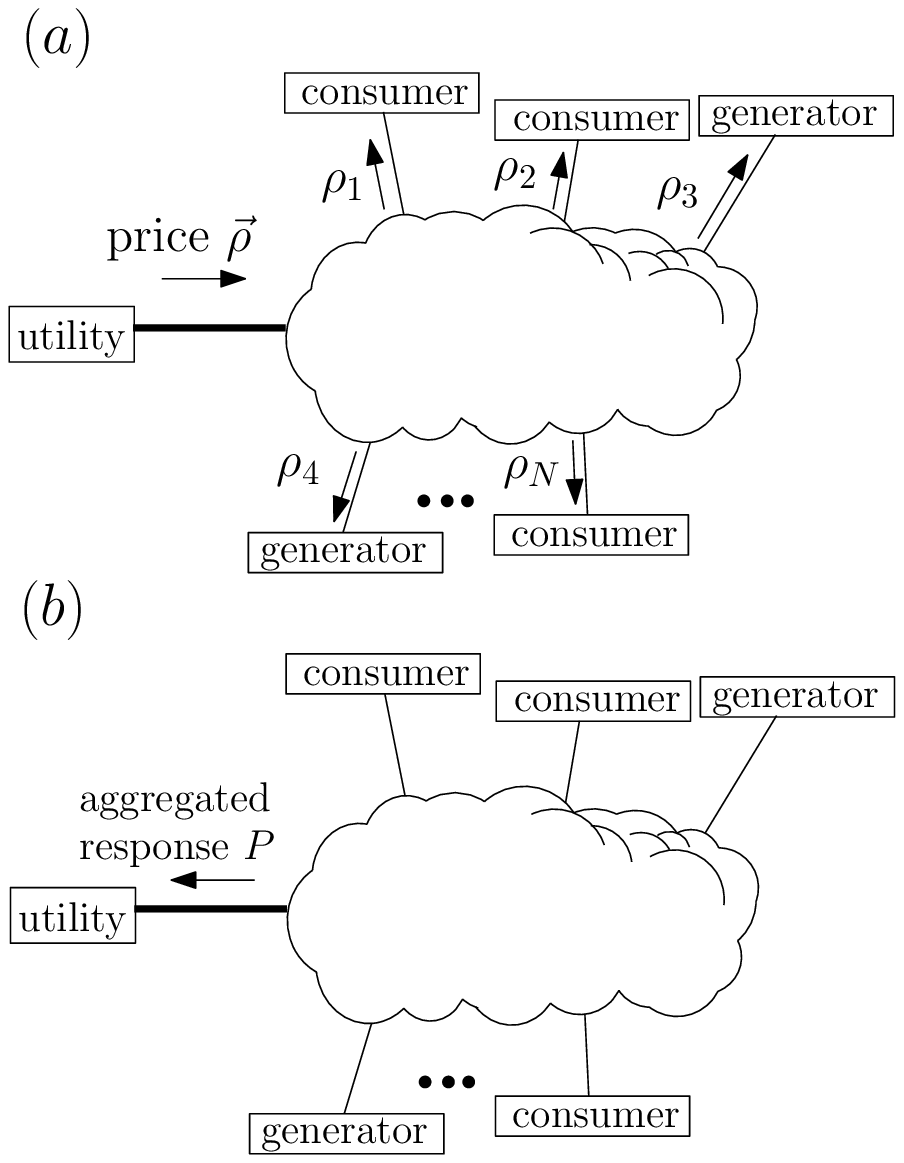}
      \caption{Scenario of the two-stage, real time, open loop control of prices and operations: 
\emph{(a)} the price signal,  including some small consumer-inhomogeneous
component, is communicated to consumers through an independent aggregating
entity; \emph{(b)} the utility senses (through electric measurements) only an
aggregated response, i.e. the cumulative/aggregated change in
consumption/production.}
      \label{fig:scenario}
   \end{figure}

We are interested in characterizing the regime where reconstruction of the
price elasticities is possible in a distribution system utilizing the multi-cast
(utility-to-consumers) communication system illustrated in
Fig.~(\ref{fig:scenario}). We analyze how the reconstruction error behaves as a
function of the Signal-to-Noise Ratio (SNR) of the aggregate power measurement and the number of available measurements per number of consumers.  For systems with small noise and
constant price elasticities, it is easy to infer the parameters optimally.
Elasticity estimation becomes significantly more difficult in very noisy
environments and when price elasticities change rapidly effectively limiting
the number of measurements available.  The problem is still solvable if one
assumes that only a small number of consumers are the ``marginal'' consumers,
i.e. only a small number of consumers respond to any particular price update.
We compare different state-of-the-art linear regression methods that
incorporate this sparsity assumption and show that their reconstruction can be
done satisfactorily given a relatively small number of samples.

In the next Section we introduce and describe our regression modeling. Section
\ref{sec:RESULTS} presents our numerical results. We conclude in Section
\ref{sec:DISC} with a discussion and future work.


\section{Regression Models for Learning Price Elasticities}
\label{sec:REGR}

We consider a distribution system consisting of $N$ individual consumers served by a single retailer/utility. We ignore losses in lines, transfer of reactive power and varying voltages, thus accounting only for redistribution of real power in a simple, capacity-based balance between production and consumption. $p_i(t)$ denotes the change in consumption of the $i$-th customer, $i=1,\dots,N$, from the previous time step $t-1$ where time is discrete, $t=1,\dots,T$. We assume the following consumer-specific, time-varying, linear relation between $p_i(t)$ and the price $\rho_i(t)$: $p_i(t)=p_i^{(0)}+\alpha_i\rho_i(t)$.  Here, $\alpha_i$ is the elasticity (linear response) rate which is under control of the customer but presumed constant for sufficiently long periods, and $p_i^{(0)}$ is the portion of the individual consumption which is insensitive to the price signal.  In this work where we only consider the open loop scenario, $\rho_i(t)$ is set by the aggregator/utility. We can model the aggregate change in consumption of the entire distribution network as the direct sum over all the consumers
\begin{align}
\label{eq:model}
P(t) & = \sum_{i=1}^{N}p_i^{(0)} + \sum_{i=1}^{N} \alpha_i\rho_i(t) + \xi(t),
\end{align}
where $\xi(t)$ is the uncertainty modeled as an aggregated zero-mean Gaussian noise with unknown variance $\beta=1/\sigma_P^2$.

Eq.~\eqref{eq:model} constitutes a standard linear regression model where the
predictors and the response variables correspond to changes in the
consumer-specific prices $\rho_i(t)$ and in the aggregated real power $P(t)$,
respectively. Our learning/reconstruction task is to estimate simultaneously
the vector of regression weights $\vec{\alpha}$ and the noise $\beta$ given the training
data $D^{\text{train}}= \{\vec{\rho}(1), P(1), \hdots, \vec{\rho}(T), P(T) \}$.
Notice that the aggregation of the  price insensitive portion of the signal,
$\sum_{i=1}^{N}p_i^{(0)}$, can be incorporated in the response vector,
therefore, without loss of generality, we can consider zero mean response
vector $P(t)$ and drop the first term from the rhs of Eq.~(\ref{eq:model}).

The Ordinary Least Squares (OLS) approach is the simplest way of solving this
linear regression problem: $\vec{\alpha}=\chi^{-1}\vec{b}$, where
$\chi$ is the input
covariance matrix, $\chi_{ij}~=~1/p\sum_{t}\rho_i(t)\rho_j(t)$, and $\vec{b}$ is
the vector of input-output covariances, $b_i=1/p\sum_{t}\rho_i(t)P(t)$. If
price elasticities $\vec{\alpha}$ do not change in time, one can obtain reliable
estimates  after a sufficiently long period of measurements.  However, either
because the individual consumptions can start affecting the price signal, or
because the individual users may change their elasticity, the periods where
$\vec{\alpha}$ remains constant can be short, limiting the small number of
samples $T$ compared to the number of consumers $N$.  In these cases, obtaining
non-biased estimates can be problematic as the typical inverse of $\chi$ is not
well defined.

One known way to address this problem is to incorporate a regularization term
into the OLS error function to penalize undesirable solutions \cite{frank93},
resulting in the following error function to minimize:
\begin{align}
\label{cost}
E(\alpha)=\frac{1}{2}\sum_{t=1}^{T}
\left(P(t)-\sum_{i=1}^{N}\alpha_i\rho_i(t)\right)^2
+\lambda\sum_{i=1}^N{|\alpha_i|}^q,
\end{align}
where $\lambda>0$ and $q\geq 0$.
Different choices of $q$ determine the prediction accuracy, interpretability of
the obtained solution (selecting variables that are relevant), and complexity
of the optimization problem. Selecting the optimal $\lambda$ is usually
performed via cross-validation.  In this work we consider three possible
choices of the penalty term in Eq.~\eqref{cost}:
\begin{itemize}
\item \textbf{Ridge regression}: \cite{Hoerl1} $q=2$. The simplest penalty term
takes the sum of squares ($\ell_2$ norm) of the weight vector $\vec{\alpha}$,
which has the effect of replacing the input covariance matrix $\chi$ with $\chi + \lambda I$,
that can be invertible.  Using ridge regression improves the prediction
accuracy, but not the interpretability of the solution.

\item \textbf{Lasso}: \cite{lasso} $q=1$. The lasso imposes an $\ell_1$ penalty
on the weights $\vec{\alpha}$ (sum of the absolute values), which has the
effect of automatically performing variable selection by setting certain
coefficients to zero and shrinking the rest.  The lasso method favors
sparse solutions while preserves the convexity (tractability) of the
optimization problem, resulting in a good compromise between prediction
accuracy, interpretability and tractability. \footnote{We use the
\texttt{glmnet} implementation for lasso in our experiments.}

\item \textbf{$\mathbf{\pmb{\ell}_0}$ norm}: $q=0$. A drawback of the lasso is
that the same $\lambda$ is used for both variable selection and shrinkage.
Consequently, lasso may select a model with too many variables to prevent
over-shrinkage of the regression coefficients \cite{relaxed}.  It is known that
using an $\ell_0$ norm instead (the number of non-zeros $\alpha_i$) improves
the selection of relevant variables, resulting in more interpretable solutions.
A complication is that for $q~<~1$, the optimization problem is non-convex and
more difficult to solve.
\end{itemize}
There are many other related regularization methods, most of them based on the
first two methods and thus resulting in convex optimization problems
(see~\cite{Friedman} for a recent account).  We restrict our analysis to the
two canonical convex methods (ridge and lasso) and a novel method for $\ell_0$
norm regularization, summarized in the next Section.

\label{sec:RESULTS}

\subsection{$\ell_0$-norm Regression}
We choose a recently introduced method \cite{vg} that performs a variational approximation on the posterior probability of the price elasticities.  It is inspired by Breiman's Garrotte \cite{breiman} and uses a spike-and-slab model \cite{spike}.

We model price elasticities $\alpha_i$ as $s_iw_i$, where the additional binary variables $s_i=\{0,1\}$ show if the customer $i$ is active ($s_i=1$) or inactive ($s_i=0$). The regression  model becomes:
\begin{align}
P(t) & = \sum_{i=1}^{N}s_iw_i\rho_i(t) + \xi(t).\notag
\end{align}
We consider the probability distribution over the parameters $(\vec{w},\vec{s},\beta)$ and compute the maximum-a-posteriori estimate from the posterior probability of the parameters given the data.  We choose the following prior distribution for $\vec{s}$:
\begin{align}
p(\vec{s}|\gamma)&=\prod_{i=1}^Np(s_i|\gamma), & p(s_i|\gamma)&=\frac{\exp(\gamma s_i)}{1+\exp(\gamma)},\notag
\end{align}
where $\gamma$ (similar to $\lambda$ before) determines the sparsity of the
solution: $\gamma\ll 0$ will favor sparse solutions and, on the contrary,
$\gamma\approx 0$ will indicate bias towards dense solutions.
The marginal posterior is approximated with the following variational bound:
\begin{align}
p(\vec{w},\beta |D,\gamma)
\propto& \sum_{\vec{s}}p(\vec{s}|\gamma) p(D|\vec{s},\vec{w},\beta)\notag\\
    \geq&
    \exp\left(-\sum_{\vec{s}}q(\vec{s})\log\frac{q(\vec{s})}
        {p(\vec{s}|\gamma)p(D|\vec{s},\vec{w},\beta)}\right),\notag
\end{align}
where we choose $q(\vec{s})=\prod_{i=1}^N(m_is_i + (1-m_i)(1-s_i))$
thus allowing us to specify $q$ with only the expected values $m_i=q_i(s_i=1)$.
For a given level of sparsity $\gamma$, the expected values $\vec{m}$ of
$\vec{s}$ and the rest of parameters $\vec{w}, \beta$ are found by iteratively
solving a set of fixed point equations defined for the expectations $m_i$, the
weights $w_i$, and the noise $\beta$.  An estimate of the price elasticity for
customer $i$ is obtained  by setting $\alpha_i' = m_iw_i$ (see \cite{vg} for more
details on the algorithm).

\section{Results}

We are only interested in testing the nontrivial case of $T<N$ because for
$T\geq N$, the elasticity of each consumer can be probed independently.  For
$T<N$, we utilize a random price strategy.  Even though the random strategy may
not be the optimal reconstruction strategy for all customer elasticity
patterns,  we expect it to be sufficiently good and robust in an average sense.
For convenience, we choose independent fluctuations for the different customers
to prevent undesired effects due to correlated predictors. In the following, we
quantitatively compare the different learning schemes introduced in Section
\ref{sec:REGR} under the aforementioned assumptions, i.e. independent random
price variations and constant customer elasticities. We analyze two simulated
scenarios: a sparse case when only $10\%$ of customers respond to the
incremental change in price and a denser case when $50\%$ of customers are
active.  The price elasticities are set to unity/zero for all active/inactive
customers. For each of the tested algorithms, parameters $\vec{\alpha}$ and
$\beta$ are estimated using a training set, $D^\text{train}$, for a
fixed hyper-parameter ($\lambda$ or $\gamma$), which
is optimized on an independent, validation set $D^\text{val}$ \cite{mitchell},
generated in the same way as $D^\text{train}$ of size $T/2$.

To compare the resulting solutions quantitatively, we compute the following
three quantities. Let $\vec{\alpha}'$ and ${\vec{\alpha}}^*$ denote the
estimated and the true price elasticities, respectively:
\begin{itemize}
\item \textbf{Generalization error}: measures how well the learning model
generalizes, i.e.  given a new vector of prices $\vec{\rho}^\text{new}$, how
the response predicted using $\vec{\alpha}'$ differs from the response obtained
using ${\vec{\alpha}}^*$. We computed it as $\sum_t
(P(t)-\sum_i{\alpha'_i\rho_i(t)})^2$, where $P(t),\vec{\rho}(t)$ belong to
$D^\text{val}$.  
\item \textbf{Area under the Receiver Operating Characteristic (ROC) curve}:
The ROC curve is calculated by thresholding the estimates $\vec{\alpha}'$.
Those $\alpha'_i$ that lie above (below) the threshold are considered as active
(inactive) customers.  For a given threshold, it is computed as the ratio
between the true positive rate and the false positive rate, where the true
positive rate means those active customers that are detected out of the actual
active ones and false positive rate means those active customers that are
detected out of the inactive ones. The ROC curve plots this relation at various
threshold settings.  The area under the curve measures the ability of the
method to correctly classify those customers that are and are not active.
A value of 1 for the area represents a perfect test whereas an 0.5 represents a
worthless test.  
\item \textbf{Reconstruction error}: measures how accurately the pattern of
price elasticities is recovered.  It is defined as  the $\ell_1$ norm of the
price elasticities differences, $\sum_i|{\alpha_i}'-{\alpha_i}^*|$.
\end{itemize}

The quality of learning depends critically on the following three dimensionless
parameters: the ratio of measurement time slots to number of samples $T/N$, the
sparsity level, and the Signal-to-Noise ratio (SNR) of the aggregate power measurement.  In the next two Subsections we consider the dependence on the number of samples and SNR.  For each
condition, we report the variations in the results over $10$ different random
instances.

\subsection{Dependence on the Number of Samples}
\label{subsec:sparse}
   \begin{figure}[t]
      \centering
      \includegraphics[scale=.5]{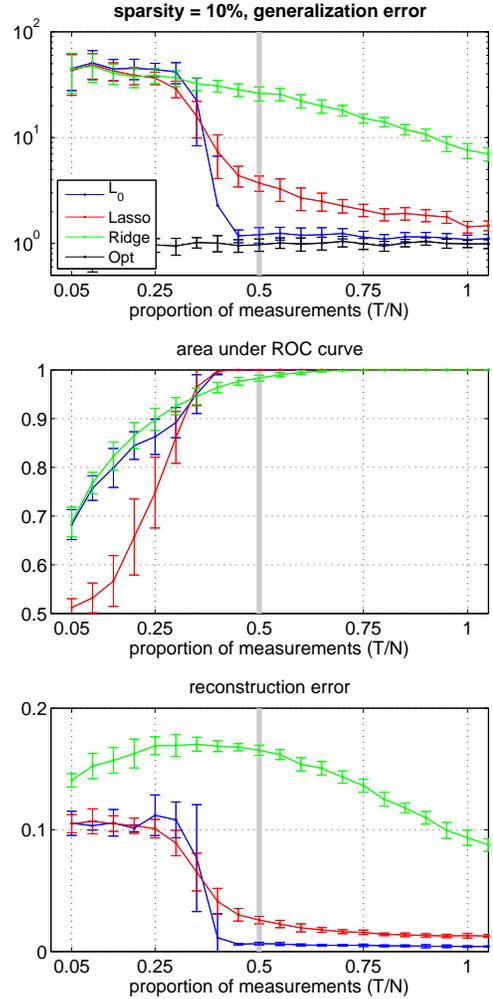}
      \caption{Results for $10\%$ of active customers vs the number of measurements.
}
      \label{fig:sparse_nsamples}
   \end{figure}

   \begin{figure}[t]
      \centering
      \includegraphics[scale=.5]{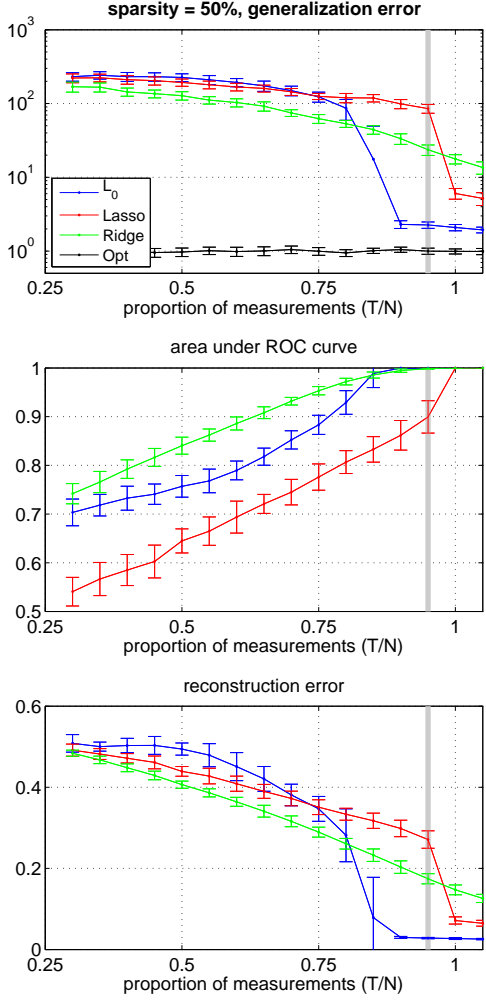}
      \caption{Results for $50\%$ of active customers vs the number of measurements.}
    \label{fig:dense_nsamples}
   \end{figure}
In our study of the dependence on $T/N$, we set the noise level to $\beta =
1/\sigma^2_P=1$. As shown in Fig.~\ref{fig:sparse_nsamples}, the
generalization errors (top plot) for the three tested methods are similar if
the number of samples is small.  Once the number of samples reaches certain
threshold (in this case $T/N\approx 40\%$) the error of $\ell_0$ drops to the
error obtained using the actual (optimal) elasticities (denoted by 'Opt' and
black curve), and the decrease in the lasso error is also significant.  On the
contrary, the performance of ridge regression improves continuously but slowly, 
remaining worse than what is shown by the other methods. The area under the ROC
curve (middle plot) shows that $\ell_0$ and ridge methods initially perform
similarly and significantly better than lasso.  This is consistent with the
fact that when the number of samples is small, the lasso outputs a trivial (all
zero) solution. However, once the threshold is reached, both lasso and $\ell_0$
outperform the ridge method.  Finally, the reconstruction error in the sparse
case (bottom plot) shows a well pronounced threshold for $\ell_0$, which
reconstructs the price elasticity pattern perfectly once $\approx40\%$ or more samples are available. The lasso error, although very small, is not totally
reduced, because some coefficients are not set to zero.  We observe that the
reconstruction error of the ridge method is not monotonic - showing an initial
increase and then decrease, which is consistent with the fact that the
ridge regression is not optimizing the reconstruction error.

The results are qualitatively different for denser problems, see
Fig.~\ref{fig:dense_nsamples}.  Testing the generalization error (top plot),
one observes an abrupt transition in both lasso and $\ell_0$ methods.  However,
the transition occurs earlier in the $\ell_0$ method ($T/N\approx 80\%$) than
in the lasso, which requires $T\approx N$ number of samples to reduce the error
significantly. Remarkably, for small $T/N$ (before the threshold) the solution
provided by the simplest method (ridge) is the best. The behavior of the area under
the ROC curves (middle plot) also differs from the sparse case -- the
performance of $\ell_0$ and lasso below the threshold is not as good as before.
Finally, the reconstruction error (bottom plot) is generally worse in this case,
and again the ridge method shows the best performace for small $T/N$.


\subsection{Dependence on the Signal-to-Noise Ratio}
   \begin{figure}[t]
      \centering
      \includegraphics[scale=.5]{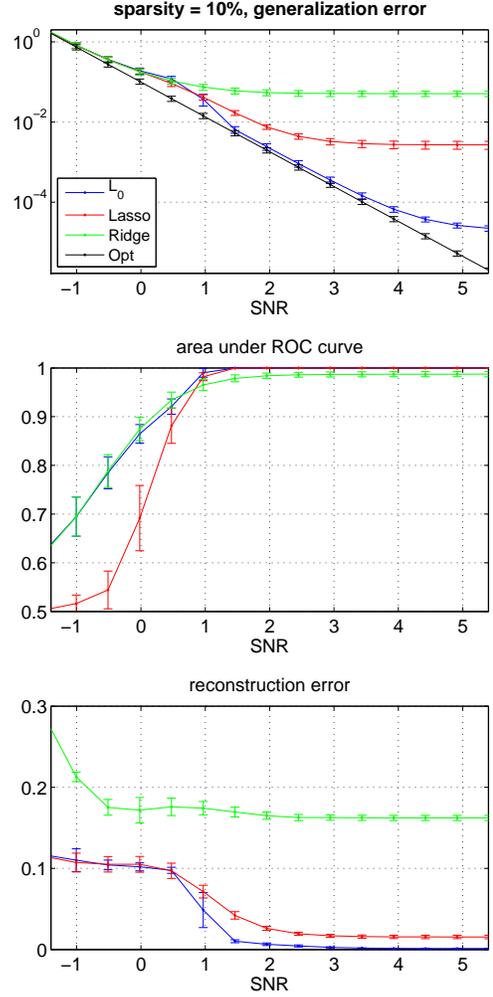}
      \caption{Results for $10\%$ of active customers vs SNR.}
      \label{fig:snr1}
   \end{figure}
   \begin{figure}[t]
      \centering
      \includegraphics[scale=.5]{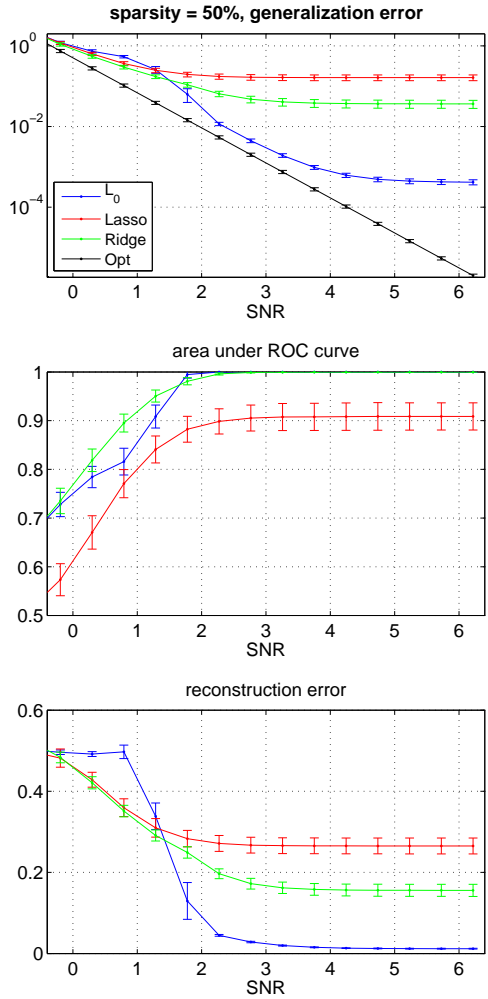}
      \caption{Results for $50\%$ of active customers vs SNR.}
      \label{fig:snr2}
      \vspace{-0.5 cm}
   \end{figure}

We now vary the SNR in a simulated environment of $N=500$ customers. We define
the SNR as the log of the average standard deviation of $\vec{\alpha}^\mathrm{T}\vec{\rho}(t)$
divided by the standard deviation $\sigma_P$. In this case, we choose the
number of time steps to be large enough to allow accurate reconstruction for
sufficiently large SNR, i.e. $T=250$ samples for a sparsity of $10\%$ and
$T=475$ samples for a sparsity of $50\%$.  These conditions are shown as gray
vertical lines in Figs.~\ref{fig:sparse_nsamples} and
\ref{fig:dense_nsamples}, respectively.

Figs.~\ref{fig:snr1} and \ref{fig:snr2} show that, at sufficiently high SNR,
$\ell_0$ performs the best. However, when the SNR is low, the other two methods
outperform $\ell_0$ in all the measures considered, but especially if the
problem is dense, see Fig.~\ref{fig:snr2}.  In the dense case, ridge regression
is the best option at low SNR. Note, however, that the bad performance of lasso
in the dense case is due to the fact that it requires more samples for denser
problems to improve over ridge, see the gray line in
Fig.~\ref{fig:dense_nsamples}.

\section{Discussion and Future Work}
\label{sec:DISC}
Our main conclusion is that the sparse reconstruction can be used to extract individual consumer price elasticities from a measured time series of aggregated consumption of real power when this aggregated power is perturbed using small, consumer-specific, random price signal variations.  For the reconstruction to be reliable, several conditions must be met: the number of time slots over which consumers do not change their elasticity should be sufficiently large,  the proportion of the consumers actually responding should be sufficiently small, and the aggregated consumption is sufficiently large so that the price-driven response is not swamped by the noise of natural fluctuations of consumption. All methods show transitions (smooth or abrupt, and sometimes at different values of the governing parameters) in reconstruction quality. In a regime where the number of samples is insufficient or when the SNR is not sufficiently large, the $\ell_0$ method performs worse than the others, and its performance degrades for denser problems.  In these bad or marginal cases, one would choose the lasso method over the $\ell_0$ method. However, when the unreliable-to-reliable transition has been crossed, the $\ell_0$ approach is preferable because it is able to reconstruct the individual price elasticities perfectly, at the cost of more computational time.  Further simulations (not discussed in the manuscript) suggest that this phase transition-like behavior becomes sharper with increase in $N$.

The technique described in this manuscript applies practically without modifications to a number of more general settings, for example to account for distributed generation (e.g. from PV systems that include local storage) sold by consumers to the utility.  This will require introducing an additional selling-price signal,  but it is otherwise identical to the description above. Generalizations accounting for other types of the exogenous signals, e.g. to outside temperature, can also be made as long as they signals are known on a consumer-specific basis.  

In a future, we will consider incorporating more details of power systems into the reconstruction, e.g.  losses, variation in voltages, and nonlinearity of power flows.  Another direction for extensions is more detailed modeling of consumer elasticity that includes the discrete and nonlinear nature of the response \cite{11TBAC}. Finally, some of the sparse reconstruction methodology discussed in this manuscript should be useful for analysis of the "closed loop" distribution markets,  e.g. the double auction markets of the Olympic Peninsula Project \cite{OlyPen} and several others discussed in recent energy market research \cite{10RDM,12CLJL,Meyn}.

\addtolength{\textheight}{-12cm}

\section*{Acknowledgement}

The work at LANL was carried out under the auspices of the National Nuclear Security Administration of the U.S. Department of Energy at Los Alamos National Laboratory under Contract No. DE-AC52-06NA25396.

\bibliographystyle{hieeetr}
\bibliography{smartgrid}

\end{document}